\documentclass[seceq]{ptptex}

\usepackage{graphicx}

\title{
Hadrons in Medium%
}

\author{
Ulrich \textsc{Mosel}$^a$, Stefan \textsc{Leupold}$^{a,c}$, Volker \textsc{Metag}$^b$%
}

\inst{
a: Institut fuer Theoretische Physik \\ b: II. Physikalisches Institut\\ Universitaet Giessen, Giessen, D-35392, Germany \\
c: Institutionen for fysik och astronomi, Uppsala Universitet, Uppsala, Sweden
}

\abst{
We review the current status of theories and experiments aiming at an understanding and a determination of the properties of light vector and scalar mesons  inside strongly interacting hadronic matter, with an emphasis on linking
calculated properties of hadrons in medium to actual observables.

}

\begin{document}

\maketitle

\section{Hadronic Spectral Functions and Observables}
\label{sec:mosel2}
In a very recent review \cite{Leupold:2009kz} we have discussed the theoretical approaches to calculate in-medium self-energies of vector and scalar mesons together with the experimental situation. In this short article, which is largely excerpted from Ref. \citen{Leupold:2009kz}, we now discuss the problem how to relate theoretically calculated  self-energies and in-medium spectral functions to actual observables. 

\subsection{Non-equilibrium effects}
All the calculations of self-energies in the literature start from the basic assumption of complete equilibrium: a hadron is embedded into nuclear matter at fixed, constant density $\rho$, without any surface effects,  and temperature $T$. This yields the self-energy $\Pi(\rho,T)$ as a function of density and temperature and this self-energy is then --- in a local-density approximation --- used to describe the hadron also in finite nuclei at higher excitation. This procedure is not without problems in the case of heavy-ion reactions where the dynamics of the collision involves also non-equilibrium phases and where the properties of the equilibrium phase change with time. This question has so far hardly been tackled (see, however, Ref.\ \citen{Schenke:2005ry}).
 On the other hand, this may be a reasonable approximation to hadron production reactions on cold, static nuclear targets where the whole reactions proceed close to equilibrium.

\subsection{Nuclear transparency}
\label{sec:nucl-transp}
The imaginary part of the in-medium self-energy $\Im \Pi_{\rm med}$ of a vector meson can be determined from nuclear transparency measurements where the total hadron yield produced in a reaction of an elementary projectile ($\gamma, \pi, p$) on a nuclear target is compared with that on a free nucleon. The nucleus then acts both as a target and as an attenuator. For the specific case of photoproduction, the total vector-meson production cross section on a nuclear target then reads in a simple Glauber approximation\cite{Muhlich:2005kf}
\begin{equation}     \label{att}
\sigma_{\gamma A} = \int d\Omega \int d^3r\, \rho(\vec{r}) \frac{d\sigma_{\gamma N}}{d\Omega}\, {\rm exp}\left(\frac{1}{|\vec{q}\,|} \int_0^{\delta r} dl \Im \Pi_{\rm med}(q,\rho(\vec{r}\,')) \right) P(\vec{r} + \delta \vec{r})
\end{equation}
with
\begin{equation}
\vec{r}\,' = \vec{r} + l \frac{\vec{r}}{r} \,, \quad \delta \vec{r} = v \frac{\gamma}{\Gamma_{\rm vac}} \frac{\vec{q}}{|\vec{q}\,|}  \,,
\end{equation}
and the (local) nucleon density $\rho(\vec{r})$; here $\Gamma_{\rm vac}$ is the free decay width of the hadron in its restframe.
Finally, $P(\vec{r})$ is the probability for the final-state hadrons to be absorbed. The imaginary part of the in-medium self-energy that determines the attenuation in (\ref{att}) is connected to the collision width of the vector meson by
\begin{equation}
\Gamma_{\rm coll} = -  \frac{1}{\omega} \Im \Pi_{\rm med} \approx  \rho \sigma v \,,
\label{Gammacoll-reit}
\end{equation}
where the last (classical) expression follows only in the low-density approximation. In all these equations $\vec{q}$ is the three-momentum of the vector meson and $\omega$ its energy. If one now assumes that the low-density limit holds, i.e.\ that the collisional width is determined by two-body collisions of the vector meson with a nucleon, then $\Gamma$ is connected to the effective in-medium two-body cross section by (\ref{Gammacoll-reit}). However, it is worthwhile to remember that already for the pion only one half of the absorption is due to two-body collisions whereas the other half involves three-body interactions.\cite{Oset:1986yi}

Eq.\ (\ref{att}) contains only absorption effects and no sidefeeding (regeneration) of the channel under study which in principle might affect the measured transparency ratios.\cite{Sibirtsev:2006yk} However, detailed comparisons with full transport calculations that contain such effects show that the Glauber approximation works very well for the cases studied here.\cite{Muhlich:2005kf,MuhlichDiss} Regeneration of meson channels through secondary interactions plays a role only for weakly absorbed mesons,
such as e.g.\ the $K^+$ (Ref.\ \citen{Effenberger:1999jc}) or low-energy pions.

If, on the other hand, the hadron being studied can decay during its propagation through the matter then also the possible reabsorption of the decay products has to be taken into account. This is a major problem if the decay is hadronic or semi-hadronic. In this case again only a good event simulation can give a reliable description of the absorption probability $P$ in (\ref{att}).

\subsection{Influence of branching ratios}
Experimental determinations of the full in-medium spectral function that involves imaginary \emph{and} real parts of the self-energy always rely on a reconstruction of the spectral function from the measured four-momenta of two decay products (e.g., $\rho \to 2\pi$; $\omega \to \pi^0 \gamma$; $\phi \to K^+K^-$; $\rho,\omega,\phi \to e^+e^-$). In a direct reaction of a microscopic probe, such as a photon, with a nucleon $N$ at a center-of-mass energy
$\sqrt{s}$ the production cross section for a vector meson $V$ with invariant mass $\mu$ is given by\cite{MuhlichDiss,Gallmeister:2007cm}
\begin{equation}
\frac{d\sigma_{\gamma N \to VN}}{d\mu} = 2 \mu \frac{1}{16\pi s|\mathbf{k}_{cm}|} |\mathcal{M}_{\gamma N \to VN}|^2 \mathcal{A}(\mu) \, |\mathbf{q}_{\rm cm}|
\end{equation}
Here $\mathcal{M}$ is the transition matrix element, $\mathbf{k}$ and $\mathbf{q}$ are the momenta of the incoming photon and the outgoing vector meson, respectively, in the cm system and $\mathcal{A}(\mu)$ is the spectral function of the vector meson\footnote{In the nuclear medium Lorentz invariance is not manifest and thus the spectral function depends both on $\mu$ and on the vector meson's three-momentum $\mathbf{q}$.}. Assuming a decay of the vector meson into two final particles $p_1$ and $p_2$ the cross section for the production of the final state, again with invariant mass $\mu$, is given by:
\begin{equation}    \label{dsigmadmufinal}
\frac{d\sigma_{\gamma N \to N(p_1,p_2)}}{d\mu} = \frac{d\sigma_{\gamma N \to VN}}{d\mu} \times \frac{\Gamma_{V \to  p_1 + p_2}}{\Gamma_{\rm tot}}(\mu) \times P_1 P_2 ~.
\end{equation}
Here $\Gamma_{\rm tot}$ is the total width of the meson $V$, obtained as a sum of the vacuum decay width, $\Gamma_{\rm vac}$, and an in-medium contribution:
\begin{equation}
\Gamma_{\rm tot} = \Gamma_{\rm vac} + \Gamma_{\rm med} ~.
\end{equation}
 The ratio $\Gamma_{V \to  p_1 + p_2}/\Gamma_{\rm tot}$ represents the branching ratio into the final state $p_1,p_2$ and $P_i$ gives the probability that the particle $p_i$ survives absorption or rescattering in the final state (we neglect here possible channel couplings). In the nuclear medium $\Gamma_{\rm tot}$ increases; it is essential that this increase is contained both in the spectral function $\mathcal{A}$ and the branching ratio.

\begin{figure}[ht]
\begin{center}
    \includegraphics[keepaspectratio,width=0.6\textwidth]{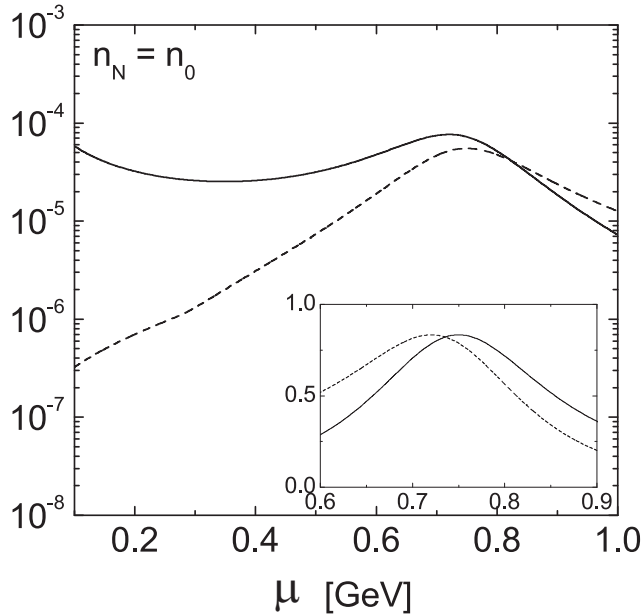}
  \end{center}
  \vspace*{8pt}
\caption{Dilepton yield from $\rho$ decay as a function of invariant mass $\mu$. The dashed line shows the $\rho$ spectral function at density $n_0$. The insert shows a magnification of the peak region on a linear scale (from Ref.\ \protect\cite{MuhlichDiss}).}
\label{rhobranch}
\end{figure}

Eq.\ (\ref{dsigmadmufinal}) shows that the invariant-mass distribution reconstructed from the four-vectors of the final particles always contains effects not only from the spectral function, but also from the branching ratio and from the final-state interactions (fsi) which also depend on $\mu$. For the case of dilepton final states the latter two are known:\cite{MuhlichDiss} the electromagnetic decay width of the $\rho$ goes like $1/\mu^3$ and can be calculated in a rather model-independent way. The branching ratio for this decay then goes like $1/(\mu^3 \Gamma_{\rm tot})$ and the dilepton fsi can be neglected. The total decay width of the $\rho$ is determined the $2\pi$ channel and possibly some collisional broadening. The $2\pi$ decay width is given by\cite{MuhlichDiss}
\begin{equation}
\Gamma_{\rho \to \pi\pi}(\mu) = \frac{f_\rho^2}{48\pi} \mu \left[1 - 4 \left(\frac{m_\pi}{\mu}\right)^2\right]^\frac{3}{2}\, \Theta(\mu - 2 m_\pi) ~.
\end{equation}
This decay width tends to go $\mu$ for $\mu \to \infty$ and to $0$ for $\mu \to 2 m_\pi$ (below that mass dilepton decay and collisional broadening are essential). Thus even when the in-medium broadening is taken into account the branching ratio becomes very large for small values of the invariant mass $\mu$ whereas in the pole region the peak of the mass distribution is slightly shifted. This is illustrated in Fig.\ \ref{rhobranch} which clearly demonstrates that the observed dilepton invariant-mass distribution has nothing to do with the actual spectral function of the $\rho$ meson.

The strongly $\mu$-dependent branching ratio shifts the observed mass distribution significantly to lower masses. For hadronic ($\pi\pi$, $KK$) or semi-hadronic ($\pi^0\gamma$) final states the mass dependence of the decay branching ratio is often not very well known and has to be modeled. While there are theoretical studies available\cite{MuhlichDiss} experimental determinations of hadronic spectral functions for the $\omega$ meson have so far not taken this into account. This is particularly critical if new particle thresholds open in the mass region of interest. This increases the total width and leads to a strong fall-off of the branching ratio with increasing mass.
For example, the branching ratio for the decay $\omega \to \pi^0 \gamma$ is strongly influenced by the opening of the $\rho \pi$ channel just in the $\omega$ mass region.\cite{MuhlichDiss} In this special case a further complication arises from the fact that the $\rho$ meson gets broadened in a medium so that as a consequence the total decay width of the $\omega$ meson may change in the nuclear medium.

\begin{figure}[ht]
\begin{center}
    \includegraphics[keepaspectratio,width=0.6\textwidth]{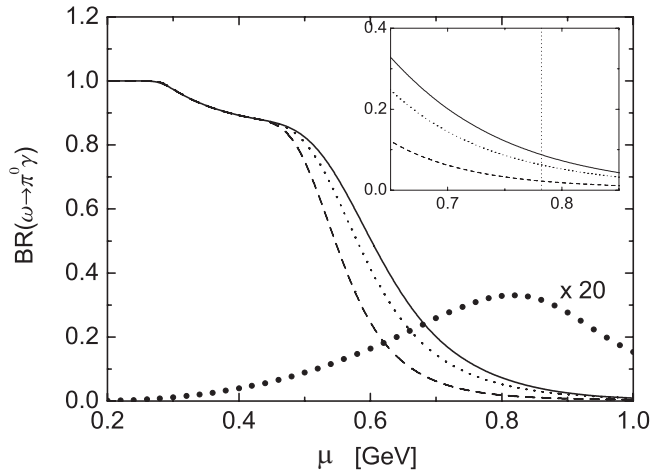}
  \end{center}
  \vspace*{8pt}
\caption{Branching ratio for $\omega \to \pi^0 \gamma$ as a function of invariant mass $\mu$. Solid: vacuum $\rho$ properties, dotted: with $\rho$ collisional broadening at $n_0$, dashed: with $\rho$ collisional broadening and dropping mass, thick dots: with collisional broadening for $\omega$ and vacuum properties for $\rho$ (from Ref.\ \protect\cite{MuhlichDiss}).}
\label{BRom}
\end{figure}

This is illustrated in Fig. \ref{BRom}
which -- due to the opening of the $\omega \to \rho \pi$ decay branch -- exhibits a strong falloff of the branching ratio just in the mass region where the $\omega$ in-medium mass could be. The figure also nicely illustrate that the in-medium properties of another meson, the $\rho$, affect the overall shape of the mass distribution for the $\omega$.

In summary, any extraction of the spectral function that relies on the determination of the invariant mass distribution $\mathcal{P}(\mu)$ from the four-momenta of the final particles
\begin{equation}
\mu = \sqrt{(p_1 + p_2)^2}
\end{equation}
requires that the partial decay width is divided out
\begin{equation}
\mathcal{A}(\mu) = \mathcal{P}(\mu) \, \frac{\Gamma_{\rm tot}}{\Gamma_{V \to  p_1 + p_2}} ~.
\end{equation}
This complication is independent of any final-state interactions that the decay products may experience.

\subsection{Observability of in-medium spectral functions}
\label{subsec:obs-broad}

Even if the mass dependence of the branching ratio is known and even if theory predicts a significant broadening of hadronic spectral functions in the nuclear medium it is not clear if this broadening can be observed.\footnote{The in-medium width which leads to this broadening can be obtained from transparency measurements as discussed in Sect. \protect\ref{sec:nucl-transp}.}
First of all, the in-medium hadrons do not all experience one given density.
Instead, because of the nuclear density profile, the relevant densities range from saturation ($\rho_0$) down to zero in the nuclear surface. Most nucleons are embedded in densities of about 1/2 to 2/3 of normal nuclear density $\rho_0$. Thus it is immediately clear that hadrons in nuclei will not exhibit properties corresponding to saturation density, but instead to a lower one. Even more important is, however, another not so straightforward effect that suppresses contributions from higher densities. We briefly outline this effect here; more details can be found in Refs.\ \cite{MuhlichDiss} and \cite{Lehr:2001ju}.

The semi-inclusive cross section for the production of final states, e.g., via a vector-meson resonance
in a photon-nucleus reaction, is obtained by
integrating Eq.\ (\ref{dsigmadmufinal}) over all nucleons. The result involves a factor
\begin{equation}
\label{inmedXsection}
\mathcal{A}(\mu)\, \frac{\Gamma_{V \to \rm final\;state}}{\Gamma_{\rm tot}}
 = \frac{1}{\pi} \frac{\mu \, \Gamma_{\rm tot}}{(\mu^2 - m_V^2)^2 + \mu^2 \Gamma^2_{\rm tot}}\, \frac{\Gamma_{V \to \rm final\;state}}{\Gamma_{\rm tot}} ~
\end{equation}
with\begin{equation}
\Gamma_{\rm tot} = \Gamma_{\rm vac} + \Gamma_{\rm med} ~.
\end{equation}
The in-medium width $\Gamma_{\rm med}$ depends on density; in the low-density approximation it is linearly proportional to the density $\rho$,
\begin{equation}
\label{Gammamed}
\Gamma_{\rm med}(\rho(r)) = \Gamma_{\rm med}(\rho_0) \, \frac{\rho(r)}{\rho_0} ~.
\end{equation}

\begin{figure}[ht]
\begin{center}
    \includegraphics[keepaspectratio,width=0.6\textwidth]{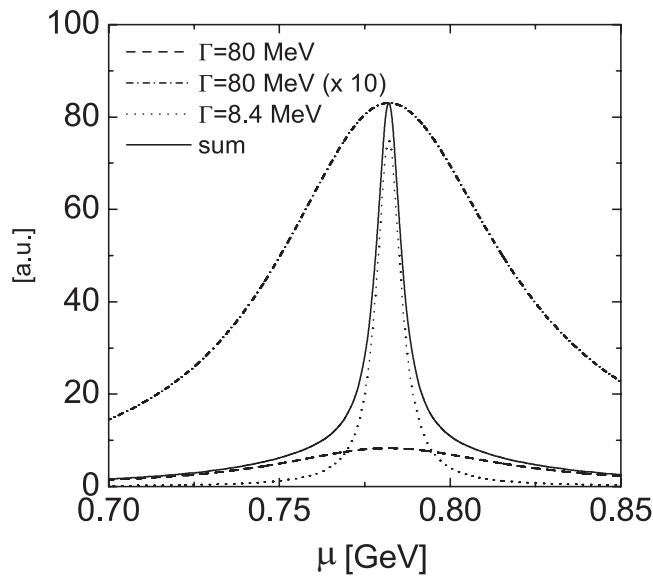}
  \end{center}
  \vspace*{8pt}
\caption{Relativistic Breit-Wigner spectral functions with equal integrated strength. Their widths are 8.4 and 80 MeV, as given in the figure. The summed spectrum has a fitted Breit-Wigner width of about 12 MeV (from Ref.\ \protect\cite{MuhlichDiss}).}
\label{om-inmedwidth}
\end{figure}

If $\Gamma_{\rm med} \gg \Gamma_{\rm vac}$, as it is the case at least for $\omega$ and $\phi$ mesons, the density dependence of the in-medium width drives the sensitivity of a meson-production experiment towards the surface. Contributions from higher densities are suppressed by order $1/\rho^2$; they are significantly broader and lower in their maximum. Inspecting Eq.\ (\ref{inmedXsection}) one sees that at the peak position, $\mu = m_V$, one suppression factor $1/\rho$ comes from the spectral function and one from the branching ratio. The effect can be seen in Fig.\ \ref{om-inmedwidth} where simply two Breit-Wigner spectral functions for the $\omega$ meson with widths differing by about a factor of 10 and equal integrated strengths have been superimposed. Since the significantly broader distribution is suppressed $\sim 1/(\Gamma_{\rm vac} + \Gamma_{\rm med})$ it changes the summed distribution only in the outer tails. In an experiment this change is difficult to separate from a background contribution.
This comparison contains only the effects of the density dependence of the spectral function. The effects of the additional density dependence of the branching ratio will only enhance the observed behavior, contributing an additional suppression factor $1/\rho$. In the example shown only the width was increased, but it is clear that the same suppression will also take place if -- in addition -- there is a shift in the peak mass that is proportional to density. The final-state interactions, not taken into account in this argument, will actually lead to an even further suppression of signals from higher densities, if the decay channel involves strongly interacting particles.

\begin{figure}[ht]
\begin{center}
    \includegraphics[keepaspectratio,width=0.7\textwidth]{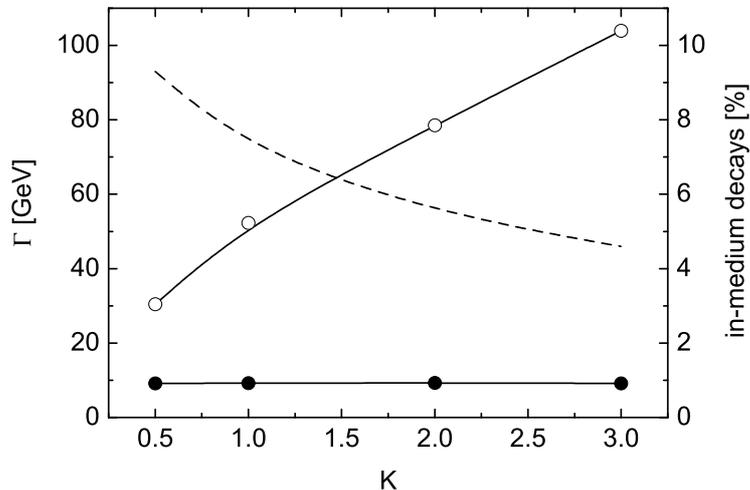}
  \end{center}
  \vspace*{8pt}
\caption{Fitted Breit-Wigner (BW) width of the $\pi^0 \gamma$ spectrum from transport simulations for the reaction $\gamma + \,^{40}$Ca $ \to \omega + X$ at 1.5 GeV photon energy. The full symbols show the BW width of the full spectrum, the open symbols that from events with in-medium decays ($\rho > 0.1 \rho_0$) only. The dashed line (right axis) shows the percentage of in-medium decays; here no cut on low momentum $\omega$ mesons has been applied. $K$ on the abscissa is a factor that multiplies the in-medium width (from Ref.\ \protect\cite{MuhlichDiss}).}
\label{om-inmedwidth1}
\end{figure}

Fig.\ \ref{om-inmedwidth1} shows the results of a full simulation for the reaction
$\gamma + \,^{40}$Ca at $E_\gamma = 1.5$ GeV.
Plotted is the width of the $\pi^0 \gamma$ spectrum as a function of a $K$ factor that multiplies the in-medium width in Eq.\ (\ref{Gammamed}) to account for a possible increase of the in-medium $\omega N$ cross section. While the total observed width stays nearly constant when $K$ is increased by a factor of 3, the width of the in-medium decay events (solid line) indeed increases with $K$, but its relative contribution (dashed line) decreases at the same time.

The suppression of higher-density contributions is less pronounced if total cross sections are considered instead of semi-inclusive or exclusive ones. In this case $\Gamma_{V \to \rm final \; state}$ in (\ref{inmedXsection}) has to be replaced by the total width $\Gamma_{\rm tot}$, effectively removing one power of $1/\rho$  so that there is less suppression for higher densities. If in addition the in-medium width is comparable to the vacuum width and not, e.g., an order of magnitude higher, then the broadening can be observable. Concerning baryon resonances, this explains why the total photoabsorption cross section on nuclei exhibits a clear broadening in the second-resonance region.\cite{Bianchi:1995vb}
A similar situation prevails in heavy-ion collisions where $\Gamma_{\rm tot}$ in the denominator of the branching ratio is canceled by $\Gamma_{\rm in}$, the width for producing the hadron in the entrance channel: While in an elementary
nuclear reaction $\Gamma_{\rm in}$ is selected by the specific experiment, the thermal
production of the considered hadron in a heavy-ion collision involves all possible
production channels, i.e.\ $\Gamma_{\rm in} \sim \Gamma_{\rm tot}$.

\subsection{Final-state interactions}
\label{sec:hadspec-obs}
In addition, the spectral functions observed in experiments by reconstructing them from hadronic decay products can be quite different from those of the original decaying meson. Final-state interactions can affect --- through rescattering --- the momenta and angles of the final-state hadrons, thus affecting also the spectral information reconstructed from the four-momenta of these hadrons. For example, the rescattering always leads to energy loss of the scattered outgoing hadron. This necessarily shifts the strength distribution towards smaller masses. In addition, hadrons can get absorbed or even reemitted (fsi); the former process will reduce the cross sections whereas the latter spoils any connection between the final observed momenta and the original spectral function.

While fsi do affect the actual observables in a significant way, if hadrons are among the decay products, their theoretical treatment is in most cases not up to the same degree of sophistication. Usually otherwise quite sophisticated in-medium calculations, which use state-of-the-art theoretical methods, do apply much more simplified methods to the treatment of fsi. Among the latter are the eikonal approximation or even a simple, absorption-only Glauber treatment. An obvious shortcoming of both of these methods is that it is ad hoc assumed that the particle that is initially hit in the very first interaction is the same as the one that ultimately leaves the nucleus on its way to the detector. However, detailed analyses have shown that there can be significant contributions from coupled-channel effects, in which a sidefeeding from an initially different channel into the final channel takes place. An example is charge transfer for pions where the more copiously produced charged pions can --- due to fsi --- be converted into uncharged ones. Such an effect plays a major role in particular in reactions with elementary probes in the incoming channel.

In addition, both the eikonal and the simple Glauber method take only flux out of a given channel; they do not yield any information on what happens with these particles. This is a major shortcoming for inclusive and semi-inclusive reactions.

An up-to-date method to treat fsi that is free of these shortcomings is provided by transport calculations. These transport calculations do take all the coupled channel effects into account, they allow for elastic and inelastic interactions and for sidefeeding and absorption. They are limited to inclusive, incoherent processes, so that exclusive particle production, for example, in coherent interactions cannot be described. However, for inclusive and semi-inclusive (or even semi-exclusive) reactions they are applicable and yield the desired results. They also provide a full dynamical simulation of the reaction and thus help to understand the reaction mechanism. State-of-the-art methods all rely on the Boltzmann-Uehling-Uhlenbeck (BUU) equation.\cite{Cassing:1990dr} A modern example of such an approach is provided by the GiBUU model.\cite{GiBUU}

\section{Summary of vector-meson experiments}

The present status of experimental results on medium modifications of vector mesons is
compiled in Table \ref{tab:exp-vec}.
\begin{table}
\centering
\begin{footnotesize}
\begin{tabular}{|c|c|c|c|c|}
\hline
 experiment & momentum & $\rho$ & $\omega $ & $\phi$\\
 & acceptance & & &  \\
\hline
\hline
 KEK-E325 & & & & \\
 pA  & $p> 0.6$ GeV/c  & $\frac{\Delta m}{m} = - 9\%$ &
 $\frac{\Delta m}{m} = - 9\%$  & $\frac{\Delta m}{m} = - 3.4\%$\\[0.3em]
 12 GeV & & $\Delta \Gamma \approx 0$ & $\Delta \Gamma \approx 0$ & $\frac{\Gamma_{\phi}(\rho_0)}{\Gamma_{\phi}}= 3.6 $\\
\hline
 CLAS & & $\Delta m \approx 0$ & & \\
 $\gamma$A  & $p> 0.8$ GeV/c & $\Delta \Gamma \approx 70$ MeV & & \\
 0.6-3.8 GeV& & $(\rho \approx \rho_0/2)$ & & \\
\hline
 CBELSA & & & $ \Delta m \approx 0 $ & \\
 /TAPS & & & $p_{\omega}< 0.5$ GeV/c & \\ \cline{4-4}
 $\gamma$A &$p>0$ MeV/c & & $ \Delta \Gamma(\rho_0) \approx 130$ MeV  &  \\
 0.9-2.2 GeV  & & & $\langle p_{\omega}\rangle = 1.1$ GeV/c & \\
\hline
 SPring8 & & & & \\
 $\gamma$A  & $p > 1.0$ GeV/c & & & $\Delta \Gamma(\rho_0) \approx 70$ MeV \\
 1.5-2.4 GeV& & & & $\langle p_{\phi}\rangle=1.8 $ GeV/c\\
\hline
CERES  & & broadening  &  &   \\
Pb+Au  & $p_t> 0$ GeV/c & favored over & & \\
158 AGeV &  & mass shift & & \\
\hline
NA60  & & $\Delta m \approx$ 0& & \\
In+In & $p_t>0$ GeV/c & strong & & \\
158 AGeV & &  broadening & & \\
\hline
\end{tabular}
\end{footnotesize}
\caption{Experimental results on in-medium modifications of $\rho$, $\omega$
  and $\phi$ mesons reported by different experiments. The reactions,
  incident energy ranges, and momentum acceptances of the detector systems are also given (adapted from Ref.\ \protect\cite{Metag:2007zza}).}
\label{tab:exp-vec}
\end{table}
The table lists the reactions, momentum ranges and
results for $\rho$,$\omega$, and $\phi$ mesons obtained in the different
experiments discussed in the review\cite{Leupold:2009kz}. A fully consistent picture has so far not been achieved.
For the $\rho$ meson, the majority of experiments observes a broadening but no
mass shift in the nuclear medium. Conflicting results reported by the KEK
experiment will have to be confirmed. Earlier claims from one experiment\cite{Trnka:2005ey}
of a dropping $\omega$ mass could not be reproduced
in a re-analysis of the data\cite{Nanova:2010sy}. A broadening of the $\omega$ meson has been
observed in elementary nuclear reactions which is in line with a depletion of
the $\omega$ yield at low momenta observed in ultra-relativistic heavy-ion
reactions but which is again in conflict with the KEK-E325 result. For the
$\phi$ meson an in-medium mass shift and a broadening has been reported.

When comparing the experimental results it should be noted
that some detector systems have no or little acceptance for low meson momenta
for which the strongest medium modifications are expected.
In view of the existing inconsistencies further experiments are needed to clarify the
situation. Corresponding experiments are planned at CLAS, HADES, PHENIX, and
the JPARC facility.

\section{Conclusions and Outlook}
The purpose of this short lecture note was a practical one: even if there were an undisputed theoretical result for the spectral function of a hadron embedded in infinite nuclear matter for an infinite time the observation of this spectral function from constructing invariant mass distributions from the decay products in an actual experiment is highly nontrivial. Not only must the final state interactions of the decay products be well known and under control, but also the mass-dependence of the decay branching ratio is entangled with the actual spectral function in the -- in principle observable - mass distribution. A further complication arises due to the density dependence of the collisional width and that of the mass-shift. If the latter two are large compared to the corresponding vacuum properties then the high-density regime gets suppressed as $1/\rho^2$  and the in-medium changes become unobservable. Strong final state interactions will tend to act in the same direction. 

\section*{Acknowledgements}

The authors gratefully acknowledge many discussions with P. Muehlich. This work has been supported by BMBF and by the LOEWE center HIC for FAIR. One of the authors (U.M.) gratefully acknowledges the hospitality by the organizers of the NFQCD10 workshop in Kyoto.


%



\begin{thebibliography}{999} 

\bibitem{Leupold:2009kz}
  S.~Leupold, V.~Metag, U.~Mosel,
  Int.\ J.\ Mod.\ Phys.\  {\bf E19 } (2010)  147-224.
  [arXiv:0907.2388 [nucl-th]].

\bibitem{Schenke:2005ry}
  B.~Schenke and C.~Greiner,
  Phys.\ Rev.\  C {\bf 73} (2006) 034909
  [arXiv:hep-ph/0509026].

\bibitem{Muhlich:2005kf}
  P.~Muhlich and U.~Mosel,
  Nucl.\ Phys.\  A {\bf 765} (2006) 188
  [arXiv:nucl-th/0510078].

\bibitem{Oset:1986yi}
  E.~Oset, Y.~Futami and H.~Toki,
  Nucl.\ Phys.\  A {\bf 448} (1986) 597.
  

\bibitem{Sibirtsev:2006yk}
  A.~Sibirtsev, H.~W.~Hammer, U.~G.~Meissner and A.~W.~Thomas,
  Eur.\ Phys.\ J.\  A {\bf 29} (2006) 209
  [arXiv:nucl-th/0606044].
  
\bibitem{MuhlichDiss}
    P.~M\"uhlich, Dissertation, Giessen University, 2007; \\
    http://www.uni-giessen.de/cms/fbz/fb07/fachgebiete/physik/einrichtungen/\\
    theorie/theorie1/publications/dissertation/muehlich\_diss/at\_download/file
  
\bibitem{Lehr:2001ju}
  J.~Lehr and U.~Mosel,
  Phys.\ Rev.\  C {\bf 64} (2001) 042202
  [arXiv:nucl-th/0105054].

\bibitem{Effenberger:1999jc}
  M.~Effenberger and U.~Mosel,
  Phys.\ Rev.\  C {\bf 62} (2000) 014605
  [arXiv:nucl-th/9908078].
  
\bibitem{Bianchi:1995vb}
  N.~Bianchi {\it et al.},
  Phys.\ Rev.\  C {\bf 54} (1996) 1688.
  


\bibitem{Gallmeister:2007cm}
  K.~Gallmeister, M.~Kaskulov, U.~Mosel and P.~Muhlich,
  Prog.\ Part.\ Nucl.\ Phys.\  {\bf 61} (2008) 283
  [arXiv:0712.2200 [nucl-th]].

\bibitem{Cassing:1990dr}
  W.~Cassing, V.~Metag, U.~Mosel and K.~Niita,
  Phys.\ Rept.\  {\bf 188} (1990) 363.

\bibitem{GiBUU}
For details on GiBUU see: http://gibuu.physik.uni-giessen.de/GiBUU

\bibitem{Metag:2007zza}
  V.~Metag,
  J.\ Phys.\ G {\bf 34} (2007) S397.
  
\bibitem{Trnka:2005ey}
  D.~Trnka {\it et al.}  [CBELSA/TAPS Collaboration],
  Phys.\ Rev.\ Lett.\  {\bf 94} (2005) 192303
  [arXiv:nucl-ex/0504010].
  

\bibitem{Nanova:2010sy}
  T.~T.~C.~/: M.~Nanova {\it et al.} [ TAPS Collaboration ],

  [arXiv:1005.5694 [nucl-ex]].






\end{thebibliography}
\end{document}